\documentclass[pra,nofootinbib,eqsecnum,floatfix]{revtex4}
\usepackage{hyperref}
\usepackage{upgreek}
\usepackage{fancyhdr}
\usepackage{lipsum}
\usepackage{amscd}
\usepackage{enumitem}
\usepackage[T1]{fontenc}
\usepackage[utf8]{inputenc}
\usepackage[usenames,dvipsnames]{xcolor}
\usepackage{graphicx,color,listings}
\usepackage{mathtools}
\usepackage{hologo}
\usepackage{mathrsfs}
\usepackage{rotating}
\usepackage{caption}
\usepackage{multirow}
\usepackage{picture}
\usepackage{eulervm}
\usepackage{graphicx}
\usepackage[nohyperlinks]{acronym}
\usepackage{bm}
\usepackage{amsmath,amssymb}
\input epsf
\numberwithin{equation}{section}

\def\TopPageMaterial{}
\def\CenterPageMaterial{}

\renewcommand\phi{\varphi}

\renewcommand{\epsilon}{\varepsilon}

\newcommand\oiint{\begingroup \displaystyle \unitlength 1pt \mkern+14.5mu\begin{picture}(0.5,2.5)\put(-2.8,2.5){\oval(12,9)}\put(6.5,1.8){\line(2,2){1.3}}\put(6.5,1.8){\line(-2,2){1.3}}\end{picture}\mkern-18.5mu\iint\endgroup}
\begin{document}
\title{Electromagnetic Classical Field Theory in a Form Independent of Specific Units}
\author{Francesco F. Summa}
\email{francesco.summa@unibas.it}
\affiliation{School of Engineering, University of Basilicata, 85100 Potenza Italy}
\begin{abstract}
\noindent
In this article we have illustrated how is possible to formulate Maxwell's equations in vacuum in an independent form of the usual systems of units. Maxwell's equations, are then specialized to the most commonly used systems of units: International system of units (SI), Gaussian normal, Gaussian rational (Heaviside-Lorentz), C.G.S. (electric), C.G.S. (magnetic), natural normal and natural rational. Both, the differential and the integral formulations of Maxwell's equations in vacuum, are illustrated. Also the covariant formulation of Maxwell's equation is illustrated.
\end{abstract}
\maketitle
\section{Introduction}
Usually, in literature and in many texts, Maxwell's equations are expressed in different systems of units. This very often leads to a great confusion and to a mixed intermediate treatment between the various systems of units. The key idea, developed in this article, is to specialize Maxwell's equations for a generic system of units, showing step by step how is possible to derive the fundamental relations of classical electrodynamics. A common strategy, to accomplish this idea, consists in the introduction of a number of unspecified constants into Maxwell's equations. For example Gelman, in his article\textsuperscript{\cite{Gelman}}, introduced five constants into Maxwell’s equations which specialize to obtain these equations in Gaussian, international system (SI), Heaviside–Lorentz (HL), C.G.S. (electric) and C.G.S. (magnetic) units. Similarly, Jackson used in the second edition of his textbook\textsuperscript{\cite{Jackson2}} the Gaussian units, while introduced four constants into Maxwell’s equations which properly specialize to yield these equations in the above-mentioned units in the third edition\textsuperscript{\cite{Jackson3}}. It is possible to demonstrate that three constants $k_{1}$, $k_{2}$ and $k_{3}$ are sufficient to express Maxwell's equations in a way independent of units. In the following table we display the values of $k_{1}$, $k_{2}$ and $k_{3}$ corresponding to International system of units (SI), Gaussian normal, Gaussian rational (Heaviside-Lorentz), natural normal, natural rational, C.G.S. (electric) and C.G.S. (magnetic).
\begin{table}[htbp]
\large
\begin{tabular}{|c|c|c|c|c|}
\hline
\multicolumn{2}{|c|}{System of Units}         & $k_{1}$          & $k_{2}$            & $k_{3}$       \\ \hline
\multicolumn{2}{|c|}{SI}                      & $k_{0}$          & $k_{m}$            & $1$           \\ \hline
\multirow{2}{*}{Gaussian} & normal            & $1$              & $\frac{1}{c}$      & $\frac{1}{c}$ \\ \cline{2-5}
                          & Heaviside-Lorentz & $\frac{1}{4\pi}$ & $\frac{1}{4\pi c}$ & $\frac{1}{c}$ \\ \hline
\multirow{2}{*}{C.G.S.}   & electric          & $1$              & $\frac{1}{c^2}$    & $1$           \\ \cline{2-5}
                          & magnetic          & $c^2$            & $1$                & $1$           \\ \hline
\multirow{2}{*}{natural \,($\hbar=c=1$)} & normal    & $1$          & $1$                & $1$            \\ \cline{2-5}
                                       & rational  & $\frac{1}{4\pi}$          & $\frac{1}{4\pi}$                & $1$            \\ \hline
\end{tabular}
\caption{The $k_{1}$,$k_{2}$ and $k_{3}$ system. The constant $k_{0}$ is defined as $k_{0}=\frac{1}{4\pi \varepsilon_{0}}$ while $k_{m}$ stands for $k_{m}=\frac{\mu_{0}}{4\pi}$}
\end{table}
\\
Using the previous table is possible to classify the different systems of units adopted in many textbooks. For example, in the context of the SI system of units we found the Vanderline's\textsuperscript{\cite{Vanderline}}, Bo Thid\'e's\textsuperscript{\cite{Bo}}, Panofsky's\textsuperscript{\cite{Panofsky}} and Griffiths\textsuperscript{\cite{Griffiths}} textbooks, while the Landau's\textsuperscript{\cite{Landau}} textbook adopts Gaussian normal units, Cohen's\textsuperscript{\cite{Cohen}} textbook adopts natural normal units and Barut's\textsuperscript{\cite{Barut}} textbook adopts natural rational units.
\section{Differential Formulation of Maxwell's Equations}
The electromagnetic field in vacuum is described by Maxwell's equations, that govern the dynamics of the electric field $\bold{E}$ and the magnetic field $\bold{B}$:
\begin{align}
\nabla\cdot\bold{E}=4 \pi k_{1}\rho\\
\nabla\times\bold{B}=4\pi k_{2}\bold{J}+(\frac{k_{2}}{k_{1}})\frac{\partial \bold{E}}{\partial t}\\
\nabla\cdot\bold{B}=0\\
\nabla\times\bold{E}=-k_{3}\frac{\partial \bold{B}}{\partial t}
\end{align}
The first and the third equations are homogeneous and correspond respectively to the Gauss law for the electric field and the magnetic one. The last two equations are the Ampere-Maxwell and the Faraday-Neumann-Lenz laws. They are not homogeneous and hold the charge density $\rho$ and the density current vector $\bold{J}$. From Maxwell's equation we extract the continuity equation. To do this, we have to consider only the inhomogeneous equations:
\begin{equation}
\begin{cases}
\nabla\cdot\bold{E}=4 \pi k_{1}{\rho}\\ \\
\nabla\times\bold{B}=4\pi k_{2}\bold{J}+(\frac{k_{2}}{k_{1}})\frac{\partial \bold{E}}{\partial t}
\end{cases}
\end{equation}
If we apply the operators $\frac{\partial}{\partial t}$ to the first equation and the $\nabla \cdot$ to the second one we obtain
\begin{equation}
\begin{cases}
\frac{\partial }{\partial t} (\nabla\cdot\bold{E})=4 \pi k_{1}\frac{\partial \rho}{\partial t}\\ \\
\nabla \cdot \left [\nabla\times\bold{B}-(\frac{k_{2}}{k_{1}})\frac{\partial \bold{E}}{\partial t}  \right ]=4\pi k_{2}(\nabla \cdot \bold{J})
\end{cases}
\end{equation}
Now we know that for a generic vector $\bold{a}$ we have $\nabla \cdot (\nabla \times \bold{a})=0$ so we obtain:
\begin{equation}
\begin{cases}
\frac{\partial }{\partial t} (\nabla\cdot\bold{E})=4 \pi k_{1}\frac{\partial \rho}{\partial t}\\ \\
-(\frac{k_{2}}{k_{1}})\nabla \cdot (\frac{\partial \bold{E}}{\partial t})  =4\pi k_{2}(\nabla \cdot \bold{J})
\end{cases}
\end{equation}
Since the derivation order is indifferent $\frac{\partial }{\partial t}\frac{\partial }{\partial x^{\alpha}}=\frac{\partial }{\partial x^{\alpha}}\frac{\partial }{\partial t}$ we have
\begin{equation}
\begin{cases}
\nabla \cdot (\frac{\partial \bold{E}}{\partial t})=4 \pi k_{1}\frac{\partial \rho}{\partial t}\\ \\
-\nabla \cdot (\frac{\partial \bold{E}}{\partial t})  =4\pi k_{1}(\nabla \cdot \bold{J})
\end{cases}
\end{equation}
If we add the two equations to each other, we obtain the continuity equation
\begin{equation}
\frac{\partial \rho}{\partial t}+\nabla\cdot\bold{J}=0
\end{equation}
that relate $\rho$ and $\bold{J}$ and expresses the law of conservation of the electric charge. The continuity equation is unit independent. The meaning of this equation became clear if we rewrite it in an integral form. Integrating both members on a volume $V^3$ we have:
\begin{equation}
\iiint_{V^3}\nabla \cdot \bold{J}\,dV^3=-\frac{\mathrm{d} }{\mathrm{d} t}\iiint_{V^3}\rho \,dV^3
\end{equation}
where, taking into account that the only quantity dependent on $t$ is the charge density $\rho$, we have taken the derivative with respect to time out of the integral sign and wrote it as a total derivative. Using now the Gauss's theorem to transform the volume integral to the first member into an integral on the closed surface $\Sigma$ which delimits $V^3$ we obtain
\begin{equation}
\oiint_{\Sigma}\bold{J}\cdot\bold{n}\,\,d\Sigma=-\frac{\mathrm{d} }{\mathrm{d} t}\iiint_{V^3} \rho \,dV^3
\end{equation}
This equation can be rewritten as
\begin{equation}
I=-\frac{dQ}{dt}
\end{equation}
where $Q$ is the total charge contained in $V^3$ and $I$ is the current that flows through $\Sigma$. If $Q$ increases there is a negative current flow, that is a certain amount of charge enters $V^3$ and vice versa. In general, in an electrodynamic problem, the Lorentz's force is introduced to account for charge particles. This force can be obtained by defining a Lorentz's density
\begin{equation}
\rho_{L}=\rho\bold{E}+k_3(\bold{J}\times \bold{B})
\end{equation}
The force exerted by the field on the entire charge distribution is given by the integral on the whole volume:
\begin{equation}
F_{L}=\iiint_{V^3} \rho_{L}\,dV^3=\iiint_{V^3}\rho\bold{E}+k_3(\bold{J}\times \bold{B})\,dV^3
\end{equation}
However, there is an important effect to be taken into consideration that we will not consider. Indeed a charged, accelerating particle, emits electromagnetic radiation which feeds back on it, affecting its motion. This effect is called a radiation reaction and can be considered negligible if the speed variation over time, therefore the acceleration, is sufficiently small. Now we want to derive the field equations for $\bold{B}$ and $\bold{E}$. To do this we derive from time the Faraday-Neumann-Lenz equation
\begin{equation}
\frac{\partial }{\partial t}(\nabla \times \bold{E})=-k_{3}\frac{\partial^2 \bold{B}}{\partial t^2}
\end{equation}
and because the derivative order is not important we can write
\begin{equation}
\nabla \times \frac{\partial \bold{E}}{\partial t}=-k_{3}\frac{\partial^2 \bold{B}}{\partial t^2}
\end{equation}
Now using the Ampere-Maxwell law we can write the following expression
\begin{equation}
\frac{\partial \bold{E}}{\partial t}=\frac{k_{1}}{k_{2}}(\nabla \times \bold{B}-4\pi k_{2}\bold{J})
\end{equation}
that have to be replaced in the previous equation to obtain
\begin{equation}
\nabla \times\left [(\frac{k_{1}}{k_{2}})(\nabla \times \bold{B})-4\pi k_{1}\bold{J}  \right ]=-k_{3}\frac{\partial^2 \bold{B}}{\partial t^2}
\end{equation}
that can be rewritten as
\begin{equation}
(\frac{k_{1}}{k_{2}})\nabla \times(\nabla \times \bold{B})-4\pi k_{1}(\nabla \times \bold{J})  =-k_{3}\frac{\partial^2 \bold{B}}{\partial t^2}
\end{equation}
After some steps and using the relation $\nabla \times (\nabla \times \bold{a})=\nabla (\nabla \cdot \bold{a})-\nabla^{2}\bold{a}$ \,for a generic vector $\bold{a}$ we can write
\begin{equation}
(\frac{k_{1}}{k_{2}})\left [ \nabla (\nabla\cdot \bold{B})-\nabla^{2}\bold{B} \right ]-4\pi k_{1}(\nabla \times \bold{J})  =-k_{3}\frac{\partial^2 \bold{B}}{\partial t^2}
\end{equation}
being \,$\nabla \cdot \bold{B}=0$\,  after some steps we obtain
\begin{equation}
-\nabla^{2}\bold{B}-4\pi k_{2}(\nabla \times \bold{J})+(\frac{k_{2}}{k_{1}}k_{3})\frac{\partial^2 \bold{B}}{\partial t^2}=0
\end{equation}
It is important to note that if we want to obtain the D'Alembert operator $\Box$ we have to define $\frac{k_{2}}{k_{1}}k_{3}=\frac{1}{c^2}$, in this way we obtain
\begin{equation}
\Box \bold{B}=-4\pi k_{2}(\nabla \times \bold{J})
\end{equation}
A similar procedure can be used to obtain the equation for the electric field $\bold{E}$
\begin{equation}
\nabla \times(\nabla \times \bold{E})=-k_{3}\nabla \times (\frac{\partial \bold{B} }{\partial t})
\end{equation}
\begin{equation}
\nabla(\nabla\cdot \bold{E})-\nabla^2 \bold{E}=-k_{3}\nabla \times (\frac{\partial \bold{B}}{\partial t})
\end{equation}
Now using the Gauss equation for the electric field $\bold{E}$ we obtain
\begin{equation}
\nabla(4 \pi k_{1} \rho)- \nabla^2 \bold{E}=-k_{3} \frac{\partial}{\partial t}(\nabla \times \bold{B})
\end{equation}
that we can write using the Ampere-Maxwell law as
\begin{equation}
4\pi k_1 \nabla \rho -\nabla^2 \bold{E}=-k_3\frac{\partial}{\partial t}(4\pi k_2 \bold{J} +\frac{k_2}{k_1}\frac{\partial \bold{E}}{\partial t})
\end{equation}
After some algebraic steps we obtain
\begin{equation}
4\pi k_1 \nabla \rho -\nabla^2 \bold{E}=-4\pi k_2 k_3\frac{\partial \bold{J}}{\partial t}-\frac{k_2}{k_1}k_3\frac{\partial^2 \bold{E}}{\partial t^2}
\end{equation}
that can be rewritten as
\begin{equation}
-\frac{k_2}{k_1}k_3\frac{\partial^2 \bold{E}}{\partial t^2}+\nabla^2 \bold{E}=4\pi k_1\nabla\rho+4\pi k_2 k_3\frac{\partial \bold{J}}{\partial t}
\end{equation}
Using the same condition as before $\frac{k_{2}}{k_{1}}k_{3}=\frac{1}{c^2}$ we can introduce the D'Alembert operator $\Box$ to obtain the following equation for the electric field $\bold{E}$
\begin{equation}
\Box \bold{E}=4\pi(k_1\nabla\rho+k_2k_3\frac{\partial \bold{J}}{\partial t})
\end{equation}
A more manageable formulation of Maxwell's equations is obtained by the introduction of a scalar potential $\phi$ and a vector potential $\bold{A}$. We know that the magnetic field $\bold{B}$ has no divergence, so there exists a function $\bold{A}(\bold{r},t)$, called vector potential, such that:
\begin{equation}
\bold{B}=\nabla\times\bold{A}
\label{potenziale vettore definito da B}
\end{equation}
If we replace this relationship in the Faraday-Neumann-Lenz law, we will note that there must exist a function called scalar potential or electrical potential such that:
\begin{equation}
\nabla\times\bold{E}+k_{3}\frac{\partial \,(\nabla\times\bold{A})}{\partial t}=0
\end{equation}
that is
\begin{equation}
\nabla\times(\bold{E}+k_{3}\frac{\partial \bold{A}}{\partial t})=0
\end{equation}
The quantity $\bold{E}+k_{3}\frac{\partial \,\bold{A}}{\partial t}$ is irrotational\textsuperscript{\cite{Helmholtz}} so we will have:
\begin{equation}
\bold{E}=-\nabla\phi-k_{3}\frac{\partial \bold{A}}{\partial t}
\label{potenziale scalare definito da E}
\end{equation}
Using the expression \ref{potenziale scalare definito da E} in the Gauss equation for the electric field we obtain:
\begin{equation}
\nabla^2 \phi+k_{3}\nabla\cdot\frac{\partial \bold{A}}{\partial t}=-4\pi k_{1}\rho
\label{equazione dei potenziali1}
\end{equation}
If we substitute the equations \ref{potenziale vettore definito da B} and \ref{potenziale scalare definito da E} in the Ampere-Maxwell law we obtain
\begin{equation}
\nabla \times (\nabla \times \bold{A})=4\pi k_{2}\bold{J}+(\frac{k_{2}}{k_{1}})\frac{\partial }{\partial t}\left [ -\nabla \phi -k_{3}\frac{\partial \bold{A}}{\partial t} \right ]
\end{equation}
Using the previous relation for a generic vector $\bold{a}$\, $\nabla \times (\nabla \times \bold{a})=\nabla (\nabla \cdot \bold{a})-\nabla^{2}\bold{a}$\, we can write
\begin{equation}
\nabla(\nabla \cdot \bold{A})-\nabla^{2}\bold{A}=4\pi k_{2}\bold{J}-(\frac{k_{2}}{k_{1}}k_{3})\frac{\partial^2 \bold{A}}{\partial t^2}-(\frac{k_{2}}{k_{1}})\nabla \frac{\partial \phi}{\partial t}
\end{equation}
that after some algebraic passages can be written as
\begin{equation}
\Box{\bold{A}}-\nabla\left [\nabla \cdot \bold{A}+(\frac{k_{2}}{k_{1}})\frac{\partial \phi}{\partial t}  \right ]+4\pi k_{2}\bold{J}=0
\label{equazione dei potenziali2}
\end{equation}
Solving Maxwell's equations is equivalent to solve the equations \ref{equazione dei potenziali1} and \ref{equazione dei potenziali2}. These equations do not uniquely define the potentials, therefore it is necessary for this purpose to introduce a gauge transformation. There are many kind of gauge transformations. The Lorentz gauge establishes the following relationship $\nabla \cdot \bold{A}+(\frac{k_{2}}{k_{1}})\frac{\partial \phi}{\partial t}=0$ while the radiation gauge establishes that $\nabla \cdot \bold{A}=0$ and $\phi=0$, the Coulomb gauge instead establishes that $\nabla \cdot \bold{A}=0$ and the temporal gauge establishes $\phi=0$.\\
\section{Integral Formulation of Maxwell equations}
Now we can see how is possible to obtain the integral version of Maxwell's equations. We pick any region $V^3$ we want and integrate both sides of each equation over that region:
\begin{align}
&\iiint_{V^3}\nabla \cdot \bold{E}\,\,dV^3=\iiint_{V^3}4\pi k_{1}\rho\,\,dV^3\\
&\iiint_{V^3}\nabla \cdot \bold{B}\,\,dV^3=0
\end{align}
On the left-hand sides we can use the Gauss's theorem, while the right sides can be  simply evaluated:
\begin{align}
&\oiint_{\Sigma}\bold{E}\cdot \bold{n}\,\,d\Sigma=4\pi k_{1}Q\\
&\oiint_{\Sigma}\bold{B}\cdot \bold{n}\,\,d\Sigma=0
\end{align}
where $Q=\sum_{i=1}^{n}q_{i}$  is the total charge contained within the region $V^3$ and $\Sigma=\partial V^3$. Gauss law tells us that the flux of the electric field out through a closed surface is (basically) equal to the charge contained inside the surface, while Gauss law for magnetism tells us that there is no such thing as a magnetic charge. For Faraday’s law we pick any surface $\Sigma$ and integrate the flux of both sides through it:
\begin{equation}
\oiint_{\Sigma}(\nabla\times\bold{E})\cdot \bold{n}\,\,d\Sigma=\oiint_{\Sigma}-k_{3}\frac{\partial \bold{B}}{\partial t}\cdot \bold{n}\,\, d\Sigma
\end{equation}
On the left we can use Stokes theorem, while on the right we can pull the derivative outside the integral:
\begin{equation}
\oint_{\partial \Sigma}\bold{E}\cdot\,ds=-k_{3}\frac{\partial }{\partial t}\Phi_{\Sigma}(\bold{B})
\end{equation}
where $\Phi_{\Sigma}(\bold{B})$ is the flux of the magnetic field $\bold{B}$ through the surface $\Sigma$. Faraday’s law tells us that a changing magnetic field induces a current around a circuit. A similar analysis helps with Ampere's law:
\begin{equation}
\nabla\times\bold{B}=4\pi k_{2}\bold{J}+(\frac{k_{2}}{k_{1}})\frac{\partial \bold{E}}{\partial t}
\end{equation}
We pick a surface and integrate:
\begin{equation}
\oiint_{\Sigma}(\nabla\times\bold{B})\cdot \bold{n}\,\,d\Sigma=\oiint_{\Sigma}4\pi k_{2}\bold{J}\cdot \bold{n}\,\, d\Sigma+\oiint_{\Sigma}(\frac{k_{2}}{k_{1}})\frac{\partial \bold{E}}{\partial t}\cdot \bold{n}\,\, d\Sigma
\end{equation}
Then we simplify each side:
\begin{equation}
\oint_{\partial \Sigma}\bold{B}\cdot ds=4\pi k_{2}I_{\Sigma}+(\frac{k_{2}}{k_{1}})\frac{\partial }{\partial t}\Phi_{\Sigma}(\bold{E})
\end{equation}
where $\Phi_{\Sigma}(\bold{E})$ is the flux of the electric field $\bold{E}$ through the surface $\Sigma$, and $I_{\Sigma}$ is the total current flowing through the surface $\Sigma$. Ampere's law tells us that a flowing current induces a magnetic field around the current, and Maxwell's correction tells us that a changing electric field behaves just like a current made of moving charges. We collect these together into the integral form of Maxwell’s equations:
\begin{align}
\oiint_{\Sigma}\bold{E}\cdot \bold{n}\,\,d\Sigma=4\pi k_{1}Q\\
\oiint_{\Sigma}\bold{B}\cdot \bold{n}\,\,d\Sigma=0\\
\oint_{\mathcal{C}}\bold{E}\cdot\,ds=-k_{3}\frac{\partial }{\partial t}\Phi_{\Sigma}(\bold{B})\\
\oint_{\mathcal{C}}\bold{B}\cdot ds=4\pi k_{2}I_{\Sigma}+(\frac{k_{2}}{k_{1}})\frac{\partial }{\partial t}\Phi_{\Sigma}(\bold{E})
\end{align}
where $\mathcal{C}=\partial \Sigma$.
\section{Energy Conservation}
We consider a system of fields and particles contained in a volume $V^3$. We can state that, if the sum of the energy associated with the electromagnetic fields in $V^3$, increases then there is a flow of electromagnetic energy from the outside to the inside and vice versa. In mathematical terms this law translates into the following equation:
\begin{equation}
\frac{\mathrm{d} U_{em}}{\mathrm{d} t}=-\Phi_{em}
\end{equation}
where $U_{em}$ is the energy of the electromagnetic field and $\Phi_{em}$ is the flow of the electromagnetic energy through the surface $\Sigma$  that contains the volume $V^3$. Introducing the energy density of the electromagnetic field $u_{em}$ and the flow of electromagnetic energy per unit of surface $\mathcal{\bold{P}}$, we will have:
\begin{align}
U_{em}=\iiint_{V^3}u_{em}\,dV^3\\
\Phi_{em}=\oiint_{\Sigma}\mathcal{\bold{P}}\cdot \bold{n}\,\,d\Sigma
\end{align}
we obtain
\begin{equation}
\frac{\mathrm{d} U_{em}}{\mathrm{d} t}=-\frac{\mathrm{d} }{\mathrm{d} t}\iiint_{V^3}u_{em}\,dV^3-\oiint_{\Sigma}\mathcal{\bold{P}}\cdot \bold{n}\,\,d\Sigma
\label{energy conservation}
\end{equation}
Now using the Gauss theorem we obtain
\begin{equation}
\frac{\mathrm{d} }{\mathrm{d} t}\iiint_{V^3}u_{em}\,dV^3=-\iiint_{V^3}\nabla\cdot \mathcal{\bold{P}} \,dV^3
\end{equation}
Being fixed the domain of integration we can bring the derivative in the sign of integral replacing it with a partial one
\begin{equation}
\iiint_{V^3}\left [\frac{\partial u_{em}}{\partial t} +\nabla\cdot\mathcal{\bold{P}}  \right ]\,dV^3=0
\end{equation}
from which we derive, given the arbitrariness of the volume, the following expression
\begin{equation}
\frac{\partial u_{em}}{\partial t} +\nabla\cdot\mathcal{\bold{P}}=0
\label{energy conservation 1}
\end{equation}
that is the continuity equation for energy. Now let's get $u_{em}$ and the Poynting vector $\mathcal{\bold{P}}$ as a function of the fields, starting from the Ampere-Maxwell and Farday-Neumann-Lenz equations
\begin{align}
\begin{cases}
\nabla\times\bold{B}=4\pi k_{2}\bold{J}+(\frac{k_{2}}{k_{1}})\frac{\partial \bold{E}}{\partial t}\\
\nabla\times\bold{E}=-k_{3}\frac{\partial \bold{B}}{\partial t}
\end{cases}
\end{align}
Multiply by scaling the first equation for $\bold{E}$ and the second for $\bold{B}$ we obtain
\begin{align}
\begin{cases}
\bold{E}\cdot(\nabla\times\bold{B})=4\pi k_{2}\bold{E}\cdot\bold{J}+(\frac{k_{2}}{k_{1}})\bold{E}\cdot\frac{\partial \bold{E}}{\partial t}\\
\bold{B}\cdot(\nabla\times\bold{E})=-k_{3}\bold{B}\cdot\frac{\partial \bold{B}}{\partial t}
\end{cases}
\end{align}
Subtracting member to member we get
\begin{equation}
\bold{E}\cdot(\nabla\times\bold{B})-\bold{B}\cdot(\nabla\times\bold{E})=4\pi k_{2}\bold{E}\cdot\bold{J}+(\frac{k_{2}}{k_{1}})\bold{E}\cdot\frac{\partial \bold{E}}{\partial t}+k_{3}\bold{B}\cdot\frac{\partial \bold{B}}{\partial t}
\end{equation}
then using vector notation $\nabla\cdot(\bold{a}\times\bold{b})=\bold{b}\cdot(\nabla\times\bold{a})-\bold{a}\cdot(\nabla\times\bold{b})$ we obtain
\begin{equation}
-\nabla \cdot(\bold{E}\times \bold{B})=4\pi k_{2}\bold{E}\cdot\bold{J}+(\frac{k_{2}}{k_{1}})\bold{E}\cdot\frac{\partial \bold{E}}{\partial t}+k_{3}\bold{B}\cdot\frac{\partial \bold{B}}{\partial t}
\end{equation}
that can be rewritten as
\begin{equation}
\frac{1}{4\pi}\left [\frac{\bold{E}}{k_1}\cdot\frac{\partial \bold{E}}{\partial t}+(\frac{k_3}{k_2})\bold{B}\cdot\frac{\partial \bold{B}}{\partial t}  \right ]+\frac{\nabla \cdot (\bold{E}\times \bold{B})}{4\pi k_2}=-\bold{J}\cdot\bold{E}
\end{equation}
Now if we note that
\begin{align}
\frac{\partial \bold{E^2}}{\partial t}=2\bold{E}\cdot\frac{\partial \bold{E}}{\partial t}\\
\frac{\partial \bold{B^2}}{\partial t}=2\bold{B}\cdot\frac{\partial \bold{B}}{\partial t}\\
\frac{\partial }{\partial t}(\bold{E^2}+\bold{B^2})=2(\bold{E}\cdot\frac{\partial \bold{E}}{\partial t}+\bold{B}\cdot\frac{\partial \bold{B}}{\partial t})
\end{align}
we can rewrite the previous equation as
\begin{equation}
\frac{1}{8\pi}\frac{\partial }{\partial t}\left [\frac{\bold{E^2}}{k_1}+(\frac{k_3}{k_2})\bold{B^2}  \right ]+\frac{\nabla \cdot (\bold{E}\times \bold{B})}{4\pi k_2}=-\bold{J}\cdot\bold{E}
\end{equation}
which compared to the equation \ref{energy conservation 1} allows us to define the energy density of the electromagnetic field and the Poynting vector as
\begin{align}
u_{em}=\frac{1}{8\pi}\left [\frac{\bold{E^2}}{k_1}+(\frac{k_3}{k_2})\bold{B^2}  \right ]\\
\mathcal{\bold{P}}=\frac{\bold{E}\times \bold{B}}{4\pi k_2}\,\,\,\,\,\,\,\,\,\,\,\,\,\,\,\,\,\,\,\,\,\,\,\,\,\,\,\,\,\,\,\,\,\,\,\,\,\,\,
\end{align}
In our case this is possible because the particles are absent so the current density $\bold{J}$ is zero.\\
\section{Momentum Conservation}
We consider the Gauss law for the electric field and the Ampere-Maxwell law:
\begin{equation}
\begin{cases}
\nabla \cdot \bold{E}=4\pi k_1 \rho \\
\nabla\times\bold{B}=4\pi k_{2}\bold{J}+(\frac{k_{2}}{k_{1}})\frac{\partial \bold{E}}{\partial t}
\end{cases}
\end{equation}
We multiply the first equation by $\bold{E}$ and the second by $\bold{B}\times$, we obtain
\begin{equation}
\begin{cases}
\bold{E}(\nabla \cdot \bold{E})=4\pi k_1 \rho \bold{E}\\
\bold{B}\times(\nabla\times\bold{B})=4\pi k_{2}(\bold{B}\times\bold{J})+(\frac{k_{2}}{k_{1}})\bold{B}\times\frac{\partial \bold{E}}{\partial t}
\end{cases}
\end{equation}
Using the vector product's anticommutative property we can rewrite the second equation as
\begin{equation}
\bold{B}\times(\nabla\times\bold{B})=-4\pi k_{2}(\bold{J}\times \bold{B})+(\frac{k_{2}}{k_{1}})\bold{B}\times\frac{\partial \bold{E}}{\partial t}
\end{equation}
Now subtracting member to member the two equations we get
\begin{equation}
\bold{E}(\nabla \cdot \bold{E})-\bold{B}\times(\nabla\times\bold{B})=4\pi k_1\rho \bold{E}+4\pi k_{2}(\bold{J}\times \bold{B})-(\frac{k_{2}}{k_{1}})\bold{B}\times\frac{\partial \bold{E}}{\partial t}
\end{equation}
which after some steps can be rewritten as
\begin{equation}
\label{mechanic momentum}
\frac{1}{4\pi}\left \{ \bold{E}(\nabla \cdot \bold{E})-\bold{B}\times(\nabla\times\bold{B})+(\frac{k_{2}}{k_{1}})\bold{B}\times\frac{\partial \bold{E}}{\partial t} \right \}= k_1\rho \bold{E}+ k_{2}(\bold{J}\times \bold{B})
\end{equation}
Now considering that
\begin{equation}
\frac{\partial }{\partial t}(\bold{E}\times \bold{B})=\bold{E}\times \frac{\partial \bold{B}}{\partial t}+\frac{\partial \bold{E}}{\partial t}\times \bold{B}
\end{equation}
we obtain
\begin{equation}
\frac{\partial \bold{E}}{\partial t}\times \bold{B}=\frac{\partial }{\partial t}(\bold{E}\times \bold{B})-\bold{E}\times \frac{\partial \bold{B}}{\partial t}
\end{equation}
that can be rewritten using the anticommutative property of vector product as
\begin{equation}
-\bold{B}\times\frac{\partial \bold{E}}{\partial t}=\frac{\partial }{\partial t}(\bold{E}\times \bold{B})-\bold{E}\times \frac{\partial \bold{B}}{\partial t}
\end{equation}
Using the previous relation and considering from Faraday's law that
\begin{equation}
\frac{\partial \bold{B}}{\partial t}=-k_{3}^{-1}(\nabla \times \bold{E})
\end{equation}
we can rewrite equation \ref{mechanic momentum} as
\begin{equation}
\small
\frac{1}{4\pi}\left \{ \bold{E}(\nabla \cdot \bold{E})-\bold{B}\times(\nabla\times\bold{B})+(\frac{k_{2}}{k_{1}})\left [ -k_{3}^{-1}\bold{E}\times(\nabla \times\bold{E})-\frac{\partial }{\partial t}(\bold{E}\times \bold{B}) \right ] \right \}= k_1\rho \bold{E}+ k_{2}(\bold{J}\times \bold{B})
\end{equation}
This relation can be integrated. Now we can define the rate of change of the particle's momentum in an electromagnetic field as
\begin{equation}
\frac{\mathrm{d} \bold{P_m}}{\mathrm{d} t}=\frac{1}{4\pi}\iiint_{V^3}\left [ k_1\rho \bold{E}+ k_{2}(\bold{J}\times \bold{B}) \right ]dV^3
\end{equation}
so we obtain
\begin{equation}
\small
\frac{\mathrm{d} \bold{P_m}}{\mathrm{d} t}=\frac{1}{4\pi}\iiint_{V^3}\left [ \bold{E}(\nabla \cdot \bold{E})-\bold{B}\times(\nabla \times
\bold{B})-(\frac{k_2}{k_1k_3})\bold{E}\times(\nabla \times \bold{E}) \right ]dV^3 -\frac{k_2}{4\pi k_1}\frac{\partial }{\partial t}\iiint_{V^3}(\bold{E}\times \bold{B})dV^3
\end{equation}
where we may identify the second integral on the right as the electromagnetic momentum in the volume $V^3$:
\begin{equation}
\frac{k_2}{4\pi k_1}\iiint_{V^3}(\bold{E}\times\bold{B})\,dV^3
\end{equation}
The integrand can be interpreted as a density of electromagnetic momentum
\begin{equation}
\Pi_{em}=\frac{k_2}{4\pi k_1}(\bold{E}\times\bold{B})
\end{equation}
We note that this momentum density is proportional to the Poynting vector $\mathcal{\bold{P}}$, with proportionality constant $\frac{k_{2}^2}{k_1}$.
\newpage
\section{Total Angular Momentum and its Decomposition}
Having defined the moment density, it is now possible to define an angular momentum density as
\begin{equation}
\bold{L_{em}}=\bold{r}\times\Pi_{em}
\end{equation}
which explicitly takes the following form
\begin{equation}
\bold{L_{em}}=\bold{r}\times\frac{k_2}{4\pi k_1}(\bold{E}\times\bold{B})
\end{equation}
Now using the relation $\bold{B}=\nabla \times \bold{A}$ we can rewrite $\bold{L_{em}}$ as
\begin{equation}
\bold{L_{em}}=\bold{r}\times\frac{k_2}{4\pi k_1}\left [ \bold{E}\times(\nabla \times \bold{A}) \right ]
\end{equation}
From the previous equation is possible to define the total angular momentum $\bold{J_{em}}$ as
\begin{equation}
\bold{J_{em}}=\iiint_{V^3}\bold{r}\times\frac{k_2}{4\pi k_1}\left [ \bold{E}\times(\nabla \times \bold{A}) \right ]\,dV^3
\end{equation}
Now if we consider the vector identity
\begin{equation}
\bold{E}\times(\nabla \times \bold{A})=\sum_{m}E^{m}\nabla A^{m}-(\bold{E}\cdot \nabla )\bold{A}
\end{equation}
we can write the total angular momentum as
\begin{equation}
\label{total angular momentum decomposition}
\bold{J_{em}}=\frac{k_2}{4\pi k_1}\sum_{m}\iiint_{V^3}E^{m}(\bold{r}\times \nabla)A^{m}\,dV^3-\frac{k_2}{4\pi k_1}\iiint_{V^3}\bold{r}\times(\bold{E}\cdot \nabla)\bold{A}\,dV^3
\end{equation}
where the first term corresponds to orbital angular momentum and the second term is to be manipulated into the form of spin angular momentum, which does not depend linearly on $\bold{r}$. Of course, the decomposition is meaningless if the gauge of $\bold{A}$ is not fixed. The gauge that is invariably chosen in this situation is the Coulomb gauge. The second term of equation \ref{total angular momentum decomposition} is treated in the following manner. We construct the vector $\bold{V}$
\begin{equation}
\bold{V}=\sum_{m}\frac{\partial }{\partial x^{m}}(E^m\bold{r}\times\bold{A})
\end{equation}
or
\begin{equation}
V|^i=\sum_{mjk}\varepsilon^{ijk}\frac{\partial }{\partial x^{m}}(E^mr^jA^k)
\end{equation}
where $\varepsilon^{ijk}$ is the Levi-Civita symbol. By using the identity
\begin{equation}
\frac{\partial }{\partial x^{m}}(E^mr^jA^k)=\delta_{jm}E^mA^k+r^j(A^k\frac{\partial E^m}{\partial x^m}+E^m\frac{\partial A^k}{\partial x^m})
\end{equation}
we get
\begin{equation}
V|^i=\sum_{jk}\varepsilon^{ijk}E^jA^k+\sum_{mjk}\varepsilon^{ijk}r^j(A^k\frac{\partial E^m}{\partial x^m}+E^m\frac{\partial A^k}{\partial x^m})
\end{equation}
or
\begin{equation}
\bold{V}=\bold{E}\times\bold{A}+\bold{r}\times\bold{A}(\nabla\cdot\bold{E})+\bold{r}\times(\bold{E}\cdot \nabla)\bold{A}
\end{equation}
Integrating over the $dV^3$ and with the assumption that
\begin{equation}
\bold{N}=\iiint_{V^3}\bold{V}\,dV^3=0
\end{equation}
we obtain
\begin{equation}
\small
\bold{J_{em}}=\frac{k_2}{4\pi k_1}\iiint_{V^3}\left [\sum_{m}E^m(\bold{r}\times \nabla)A^m +\bold{E}\times\bold{A}+\bold{r}\times\bold{A}(\nabla\cdot\bold{E} ) \right ]dV^3=\bold{L}+\bold{S}+\frac{k_2}{4\pi k_1}\iiint_{V^3}\bold{r}\times\bold{A}(\nabla\cdot\bold{E} )\,dV^3
\end{equation}
The first term $\bold{L}$ is denoted the orbital term, the second term $\bold{S}$ is the spin term and the third term is non-zero only in the presence of charge.
\section{Covariant formulation of Maxwell's equations}
In this section we adopt the Einstein's summation convention on repeated indices. To express a covariant formulation of Maxwell's equations is necessary to introduce the concept of four-vector and the metric signature of the Minkowski spacetime $g_{\mu v}$. A four-vector in spacetime can be represented in the relativistic notation as $\bold{F}_{\mu}=(f_0,\bold{F})$, where $f_0$ is its time component and $\bold{F}$ is space component. The metric signature of the Minkowski spacetime is
\begin{equation}
g_{\mu v}=g^{\mu v}=\begin{bmatrix}
1 &0  &0  &0 \\
 0&  -1&0  &0 \\
 0&  0&  -1& 0\\
0 &  0&  0& -1
\end{bmatrix}
\end{equation}
Derivatives in spacetime are defined by $\partial_{\mu}=\frac{\partial }{\partial x^{\mu}}=\left [ \frac{1}{c}\frac{\partial }{\partial t},\nabla \right ]$ and $\partial^{\mu}=\frac{\partial }{\partial x_{\mu}}=\left [ \frac{1}{c}\frac{\partial }{\partial t},-\nabla \right ]$. The source of the electromagnetic field tensor $\bold{F}^{\mu v}$ is the four-current
\begin{equation}
\bold{J}^{\mu}=(c\rho,\bold{J})
\end{equation}
The electromagnetic field tensor $\bold{F}^{\mu v}$ satisfies the Maxwell's equations in $k_1$,$k_2$ and $k_3$ units:
\begin{align}
&\partial_{\gamma}\bold{F}_{\mu v}+\partial_{\mu}\bold{F}_{v\gamma}+\partial_{v}\bold{F}_{\gamma\mu}=0\\
\label{electromagnetic lagrangian density}
&\partial_{\mu}\bold{F}^{\mu v}=4\pi k_2 \bold{J}^{v}
\end{align}
where the first equation stands for the homogeneous Maxwell's equations and come from the definition of an antisymmetric tensor, while the second must derive from the Lagrangian density of the electromagnetic field and refers to the non-homogeneous equations. We have to note that the covariant expression of the homogeneous Maxwell's equations is more simple if we introduce the concept of the dual of $\bold{F}^{\mu v}$, defined as $\bold{F^{*}}^{\mu v}=\frac{1}{2}\varepsilon^{\mu v k \sigma}\bold{F}_{k \sigma}$. Using the definition of $\bold{F^{*}}^{\mu v}$ the homogeneous Maxwell's equations can be written as $\partial_{\mu}\bold{F^{*}}^{\mu v}=0$. The tensor $\bold{F}^{\mu v}$ is defined as
\begin{equation}
\bold{F}^{\mu v} =\begin{bmatrix}
0 &-\frac{k_2}{k_1}cE_{x}  &-\frac{k_2}{k_1}cE_{y}  &-\frac{k_2}{k_1}cE_{z} \\
\frac{k_2}{k_1}cE_{x} &  0&-B_{z}  &B_{y} \\
\frac{k_2}{k_1}cE_{y} &B_{z}  &  0&-B_{x} \\
\frac{k_2}{k_1}cE_{z} &-B_{y}  &B_{x}  & 0
\end{bmatrix}
\end{equation}
We define a vector polar if the sign of its components changes if we reverse the direction of the Cartesian axes while axial a vector that does not enjoy this property. Using this definitions we call $F^{i0}$ and $F^{ij}$ the polar and the axial components of $\bold{F}^{\mu v}$ defined respectively as
\begin{align}
&F^{i0}=\frac{k_2}{k_1}c(\bold{E})^i\\
&F^{ij}=-\varepsilon^{ijk}(\bold{B})_k
\end{align}
with $\varepsilon^{ijk}$ the Levi-Civita symbol, $(\bold{E})^i$ and $(\bold{B})_k$ that represent the components of the electric and magnetic fields. The components of the dual tensor $\bold{F^{*}}^{\mu v}$ can be obtained from those of $\bold{F}^{\mu v}$ by making the following changes:
\begin{align}
&{F^*}^{i0}\rightarrow (\bold{B})^i \\
&{F^*}^{ij}\rightarrow \frac{1}{k_3 c}\varepsilon^{ijk}(\bold{E})_k
\end{align}
where we have used the relation $\frac{k_2}{k_1}c=\frac{1}{k_3 c}$. With the aid of the above definition, we can write the following four-vectors in the $(1+3)$ notation as:
\begin{align}
&\partial_{\mu}\bold{F}^{\mu v}=(\frac{k_2}{k_1}c\nabla\cdot \bold{E},\nabla \times \bold{B}-\frac{k_2}{k_1}\frac{\partial \bold{E}}{\partial t})\\
&\partial_{\mu}\bold{F^{*}}^{\mu v}=(\nabla \cdot \bold{B},-\nabla \times \bold{E}-k_3\frac{\partial \bold{B}}{\partial t})
\end{align}
The four-potential is defined in the $(1+3)$ notation as
\begin{equation}
\bold{A}^{\mu}=(\frac{k_2}{k_1}c\phi,\bold{A})
\end{equation}
In order for the equation \ref{electromagnetic lagrangian density} to be verified, the Lagrangian density must have the following form
\begin{equation}
\mathscr{L}=-\frac{k_3}{16\pi k_2}\bold{F}^{\mu v}\bold{F}_{\mu v}-k_3\bold{J}_{\mu}\bold{A}^{\mu}
\end{equation}
that can be rewritten as
\begin{equation}
\mathscr{L}=\frac{k_3}{16\pi k_2}(\partial^{\mu}\bold{A}^{v}-\partial^{v}\bold{A}^{\mu})(\partial_{\mu}\bold{A}_{v}-\partial_{v}\bold{A}_{\mu})-k_3\bold{J}_{\mu}\bold{A}^{\mu}
\end{equation}
or in non manifest covariant notation as
\begin{equation}
\mathscr{L}=\frac{1}{8\pi}\left [ \frac{\left | \bold{E}^2 \right |}{k_1}-\frac{k_3}{k_2}\left | \bold{B}^2 \right | \right ]-\rho \phi +k_3\bold{J}\cdot \bold{A}
\label{lagrangian density non manifest covariant}
\end{equation}
To show that starting from the Lagrangian density, defined above, which is manifestly covariant or not, only two of the four Maxwell's equations are obtained, we apply the Euler Lagrange fields equations, defined for a generic field $\psi_j$ as
\begin{equation}
\sum_{j}\left [\frac{\partial \mathscr{L}}{\partial \psi_j}-\frac{\partial }{\partial t}( \frac{\partial \mathscr{L}}{\partial( \frac{\partial \psi_j}{\partial t}) })-\sum_{k=1}^{3}\frac{\partial }{\partial x_k}( \frac{\partial \mathscr{L}}{\partial( \frac{\partial \psi_j}{\partial x_k}) })\right ] =0
\label{Euler Lagrange fields equations}
\end{equation}
to the Lagrangian density defined in equation \ref{lagrangian density non manifest covariant}. To do this we have to redefine the equation \ref{lagrangian density non manifest covariant}, by introducing the expressions for $\bold{E}$ and $\bold{B}$, as  function of potentials, by doing so we get
\begin{equation}
\mathscr{L}=\frac{1}{8\pi}\left [ \frac{(\nabla \phi )^2}{k_1}+\frac{2 k_3 \nabla \phi }{k_1}\frac{\partial \bold{A}}{\partial t}+\frac{k_{3}^2}{k_1}(\frac{\partial \bold{A}}{\partial t})^2 -\frac{k_3}{k_2}(\nabla \times \bold{A})^2 \right ]-\rho \phi +k_3\bold{J}\cdot\bold{A}
\end{equation}
In this context the fields defined in the equations \ref{Euler Lagrange fields equations} are $\phi$ and $\bold{A}$. Applying Euler Lagrange's equations to $\phi$ we get
\begin{align}
&\frac{\partial \mathscr{L}}{\partial (\frac{\partial \phi}{\partial x_k})}=-\frac{E_k}{4\pi k_1}\\
&\frac{\partial \mathscr{L}}{\partial (\frac{\partial \phi}{\partial t})}=0\\
&\frac{\partial \mathscr{L}}{\partial \phi}=-\rho
\end{align}
from which we obtain
\begin{equation}
\nabla\cdot\bold{E}=4 \pi k_{1}\rho
\end{equation}
We now carry out a similar procedure for $\bold{A}$. For sake of simplicity we consider only the component $A_x$, we get
\begin{align}
&\frac{\partial \mathscr{L}}{\partial (\frac{\partial A_x}{\partial t})}=-\frac{k_3}{4\pi k_1}E_x\\
&\frac{\partial \mathscr{L}}{\partial (\frac{\partial A_x}{\partial x_2})}=\frac{k_3}{4\pi k_2}B_z\\
&\frac{\partial \mathscr{L}}{\partial (\frac{\partial A_x}{\partial x_3})}=-\frac{k_3}{4\pi k_2}B_y\\
&\frac{\partial \mathscr{L}}{\partial (\frac{\partial A_x}{\partial x_1})}=0\\
&\frac{\partial \mathscr{L}}{\partial A_x}=k_3J_x
\end{align}
from which we derive
\begin{equation}
\frac{k_3}{4\pi k_2}(\frac{\partial B_z}{\partial x_2}-\frac{\partial B_y}{\partial x_3})-\frac{k_3}{4\pi k_1}\frac{\partial E_x}{\partial t}-k_3J_x=0
\end{equation}
that is the first component of the equation
\begin{equation}
\nabla\times\bold{B}=4\pi k_{2}\bold{J}+(\frac{k_{2}}{k_{1}})\frac{\partial \bold{E}}{\partial t}
\end{equation}
From the previous relations it is easy to derive the Hamiltonian density of our system
\begin{equation}
\mathscr{H}=\sum_j \pi_j \frac{\partial \psi_j}{\partial t}- \mathscr{L}
\end{equation}
considering that the canonical momentum densities $\pi_j$ are defined as
\begin{equation}
\pi_j=\frac{\partial \mathscr{L}}{\partial (\frac{\partial \psi_j}{\partial t})}
\end{equation}
Indeed we have
\begin{align}
&\pi_{A}=\frac{\partial \mathscr{L}}{\partial (\frac{\partial \bold{A}}{\partial t})}=-\frac{k_3}{4\pi k_1}\bold{E}=\frac{k_3}{4\pi k_1}\left [ \nabla\phi+k_3\frac{\partial \bold{A}}{\partial t} \right ]\\
&\pi_{\phi}=\frac{\partial \mathscr{L}}{\partial (\frac{\partial \phi}{\partial t})}=0
\end{align}
from which we get
\begin{equation}
\mathscr{H}=\frac{1}{8\pi}\left [ \frac{k_{3}^2}{k_1}(\frac{\partial \bold{A}}{\partial t})^2 +\frac{k_3}{k_2}(\nabla \times \bold{A})^2 -\frac{(\nabla \phi)^2}{k_1} \right ]+\rho \phi-k_3\bold{J}\cdot \bold{A}
\end{equation}
If we want to obtain all Maxwell's equations by apply the Euler Lagrange fields equations, we have to define a new Lagrangian density. This theoretical problem is mentioned in the Baker's article\textsuperscript{\cite{Baker}} and it is not the aim of this paper.
\section{Discussion}
Graduate students actually work with equations in both SI and Gaussian units, which does not seem to be the best alternative from a pedagogical point of view. For this reason the  $k_1\,k_2\,k_3$ system may be a pedagogical alternative for undergraduate and graduate students who can solve electromagnetic problems without having to work in a specific system of units.
\section{Conclusions}
In this paper we have considered the old idea of writing electromagnetic equations in a form independent of specific units showing step by step how it is possible to derive the fundamental relations of classical electrodynamics. We have followed the work of Jackson\textsuperscript{\cite{Jackson3}}, that introduce four constants in Maxwell's equations, showing that only three are needed. We have shown that for each system of units the relationship $\frac{k_{2}}{k_{1}}k_{3}=\frac{1}{c^2}$ must occur to define the D'Alembert operator. This paper can be used as an introductive chapter in many courses in order to clarify all the mathematical derivations, performed step by step, that lead to the different formulations of the Classical Electromagnetic Field Theory, also in the covariant notation adopted in Special Relativity.

\end{document}